\documentclass[a4paper]{ESASPCS13Style}
\usepackage{epsfig}

\begin{document}

\title{Near-Infrared Photometry and Spectroscopy of L and T Dwarfs:
  The Effects of Clouds, Gravity, and Effective Temperature}

\author{
        D.~A.~Golimowski\inst{1}
	\and
        S.~K.~Leggett\inst{2}
	\and
        M.~S.~Marley\inst{3}
	\and
        X.~Fan\inst{4}
	\and
        T.~R.~Geballe\inst{5}
	\and
        G.~R.~Knapp\inst{6}
        }
\institute{
        Department of Physics and Astronomy, 
        Johns Hopkins University, 
        3400 N.~Charles St., 
        Baltimore, MD 21218
   	\and
        United Kingdom Infrared Telescope,
 	Joint Astronomy Centre,
	660 N.~A'ohoku Place,
	Hilo, HI 96720
	\and
	NASA-Ames Research Center,
	Mail Stop 245-3,
	Moffett Field, CA 94035
	\and
	Steward Observatory,
	University of Arizona,
	Tucson, AZ 85721
	\and
	Gemini Observatory,
	670 N.~A'ohoku Place,
	Hilo, HI 96720
	\and
	Princeton University Observatory,
	Princeton, NJ 08544
        }

\maketitle 

\begin{abstract}

We present new $JHKL'M'$ photometry on the MKO system for a large sample
of L and T dwarfs identified from SDSS and 2MASS and classified
according to the scheme of \cite*{geb02}.  We have compiled a sample
of 105 L and T dwarfs that are uniformly classified and measured
on a single photometric system.  The scattered $JHK$ spectral indices and
colors of L dwarfs are likely caused by variations in the altitudes,
distributions, and thicknesses of condensate clouds. Scatter in the $H$--$K$
colors of late T dwarfs probably reflects the sensitivity of the $K$-band
flux to pressure induced H$_2$ opacity, which itself is sensitive to
surface gravity.  The $M'$ luminosities of late-T dwarfs are 1.5--2.5
times fainter than predicted under conditions of chemical equilibrium.
We have computed $L_{\rm bol}$ and $T_{\rm eff}$ for 42 L and T dwarfs whose
trigonometric parallaxes have been measured. We find that $T_{\rm eff} \approx 
1450$~K for types L7--T4, which supports recent models that attribute the
changing $JHK$-band luminosities and spectral features across the L--T
transition to rapid changes in the condensate clouds over a narrow
range of $T_{\rm eff}$. We compute $T_{\rm eff} = 600$--750~K for 
2MASS~0415-0935 (T9), which supplants Gl~570D as the coolest known brown 
dwarf.

\end{abstract}

\section{Photometric Data}
\label{sec:photdata}

\begin{figure}[!t]
  \begin{center}
    \epsfig{file=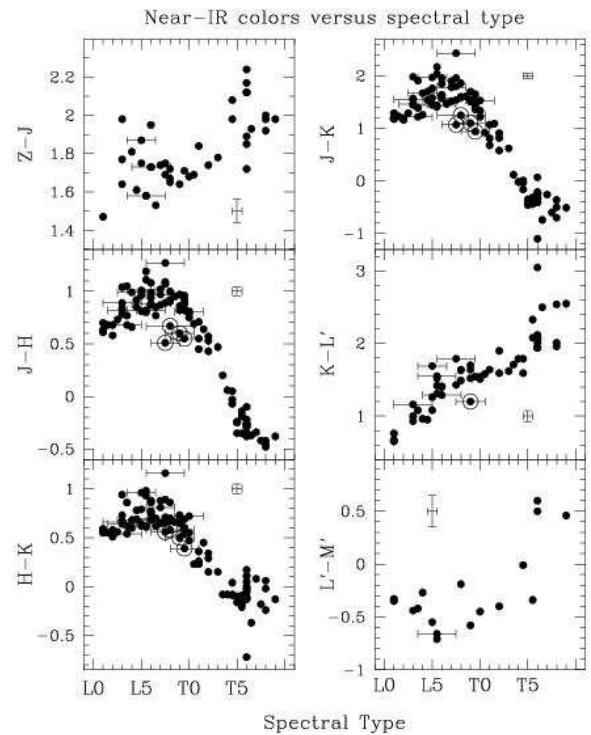, width=10cm}
  \end{center}
\caption{Near-IR colors of L and T dwarfs.  Typical error bars are shown.  Encircled
points represent the unusually blue late-L dwarfs SDSS~J0805+4812, SDSS~J0931+0327, SDSS~J1104+5548,
and SDSS~J1331--0116.
\label{dgolimowskif1}}
\end{figure}

We have compiled complete or partial 1--5~$\mu$m photometry for 105 
spectroscopically-confirmed L and T dwarfs using United Kingdom Infrared
Telescope (UKIRT) imaging cameras and UKIRT $Z$ and Mauna Kea Observatory
(MKO) $JHKL'M'$ filters.  Diagrams of color versus near-infrared spectral
type (\cite{geb02}) are shown in Figure~\ref{dgolimowskif1} for various 
combinations of bandpasses.

The $Z$ through $K$ colors of L dwarfs are scattered because their 1--$2.5~\mu$m
``photospheres'' coincide with clouds of Fe and silicate condensates with varying
optical depths (\cite{ack01}, \cite{mar02}, \cite{tsu02}).  Thus, near-IR colors 
are not good indicators of L subtype.  On the other hand, $J$--$H$ is a good 
indicator of T subtype presumably because, for $T_{\rm eff} < 1400$~K, the 
condensate clouds have settled below the near-infrared photosphere.  Conversely,
$H$--$K$ is broadly scattered among T dwarfs because the $K$ flux is affected by
pressure-induced H$_2$ absorption, which is very sensitive to surface gravity
(\cite{bur02}, and references therein).  The $Z$--$J$ colors of T dwarfs are also 
scattered, possibly because of variable process(es) responsible for removing
condensates at the L--T transition, or because the $Z$ flux is affected by the
gravity- and metallicity-sensitive wings of the Na-D and K resonance lines.

The $K$--$L'$ and $L'$--$M'$ colors of L6--T4 dwarfs are nearly constant.  The constancy
of $K$--$L'$ over this range of types is explained by balanced CH$_4$ absorption in the 
$K$ and $L'$ bands and the redistribution of continuum flux as the condensate clouds 
settle.  On the other hand, $L'$--$M'$ is largely unaffected by the settling clouds.  
Its constancy for types L6--T4 reflects the constancy of $T_{\rm eff}$ throughout the 
L--T transition. (See \S3.)

\section{Spectroscopic Data}
\label{sec:specdata}

We have obtained $JHK$ spectra of 23 new L dwarfs and 14 new T0--T7 dwarfs identified 
from the Sloan Digital Sky Survey (SDSS) photometric database using UKIRT's CGS4 and 
UIST spectrographs.  We have also obtained new spectra of 11 L and T dwarfs previously 
identified from the Two-Micron All Sky Survey (2MASS), including the coolest known T 
dwarf, 2MASS~J0415--0935.  We have classified these dwarfs according to the scheme of 
\cite*{geb02}.  Our augmented sample shows that some spectral indices of this scheme 
are not internally consistent for L dwarfs because of their apparent sensitivity to 
cloud optical depth.  Some adjustment of these indices is warranted.

\begin{figure}[!b]
  \begin{center}
    \epsfig{file=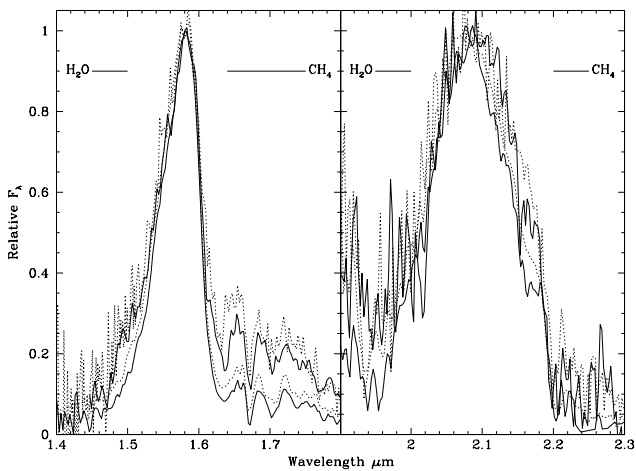, angle=-90, width=9cm}
  \end{center}
\caption{Normalized $H$ and $K$ spectra ($R \approx 600$) for {\it (top to bottom)}
SDSS~J1624+0029 (T6), SDSS~J1758+4633 (T7), Gl~570D (T8), and 2MASS~J0415--0935 (T9).
\label{dgolimowskif2}}
\end{figure}

Figure~\ref{dgolimowskif2} shows the $H$ and $K$ spectra of representative T6, T7, T8, and T9 
dwarfs.  These spectra show a steady increase in the depths of the H$_2$O and CH$_4$
bands with increasing spectral type.  The relatively small spectral differences between 
the dwarfs designated T8 and T9 may be more indicative of a half-subtype increment, rather 
than a whole-subtype increment.  Regardless, this sequence suggests that there is room for 
one more subtype, which would have essentially zero flux at 1.45, 1.7, and 2.25~$\mu$m.

\begin{figure}[!t]
  \begin{center}
    \epsfig{file=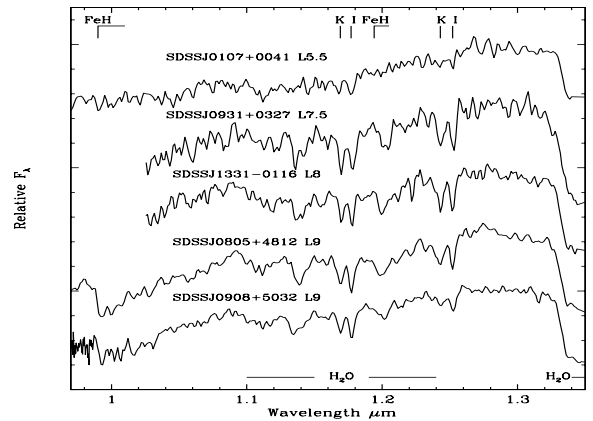, height=6cm, width=8cm}
  \end{center}
\caption{Normalized $J$ spectra ($R \approx 600$) for 3 of the 4 unusually blue late-L dwarfs,
bracketed by more typical L5.5 and L9 dwarfs.  Strong FeH, K~I, and H$_2$O absorption
features are identified.
\label{dgolimowskif3}}
\end{figure}

The effects of varying cloud optical depth and gravity are manifested not only in the 
broadband colors of L and T dwarfs (\S1), but in the equivalent widths of narrower
atomic and molecular absorption lines.  Figure~\ref{dgolimowskif3} compares the $J$ spectra 
of three unusually blue late-L dwarfs (marked with circles in Figure~\ref{dgolimowskif1}) with
those of more typical L dwarfs.  The enhanced strengths of the FeH and K~I features in 
the former group mimic those of earlier L dwarfs, which suggests that either the condensate 
clouds of these late-L dwarfs are unusually thin, or these dwarfs have low metallicity.

\begin{figure}[!b]
  \begin{center}
    \epsfig{file=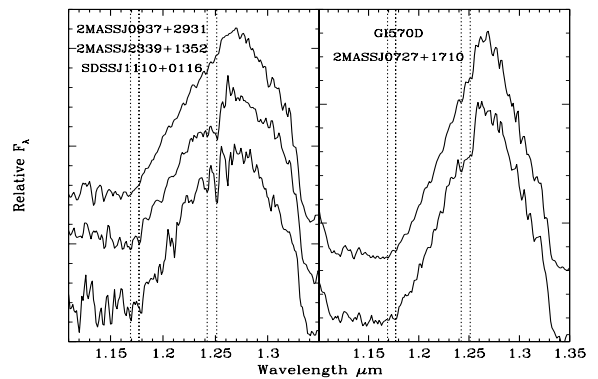, height=5.5cm, width=8cm}
  \end{center}
\caption{Normalized $J$ spectra ($R \approx 600$) for T6 {\it (left)} and T8 {\it (right)} dwarfs. 
The K~I absorption strengths and the $H$--$K$ colors of these dwarfs increase from top to bottom.
Both trends indicate decreasing surface gravity.
\label{dgolimowskif4}}
\end{figure}

Figure~\ref{dgolimowskif4} shows $J$ spectra of several T6 and T8 dwarfs that exhibit progressively
stronger absorption by K~I at 1.243 and 1.254~$\mu$m.  Recent investigations of spectroscopic
gravity indicators in T dwarfs indicate that this doublet strengthens with decreasing gravity
(\cite{mar03}).  This trend is supported by the increasing $H$--$K$ colors of these dwarfs (see
Figure~\ref{dgolimowskif7}), which reflect progressively less pressure-induced H$_2$ absorption of $K$-band 
flux as gravity decreases
(\cite{bur02}).

\section{Bolometric Corrections and $T_{\rm eff}$}

We used our $ZJHK$ spectra, our $ZJHKL'$ photometry, and recently published parallaxes (\cite{dah02},
\cite{tin03}, \cite{vrb04}) to compute $BC_K$ and $T_{\rm eff}$ for late-M, L, and T dwarfs.  In doing
so, we assumed a Rayleigh-Jeans flux distribution longward of $L'$ appropriately corrected for CH$_4$ 
and CO absorption between 3 and $5~\mu$m.  We used available $M'$ photometry to confirm the accuracy
of these assumptions.  To compute $T_{\rm eff}$, we adopted the ranges of brown-dwarf radii predicted 
by the evolutionary models of \cite*{bur97}, \cite*{bar98}, and \cite*{cha00} for ages 0.1--10~Gyr.
(Narrower ranges were used for those dwarfs whose ages have been spectroscopically constrained.)
Nominal values of $T_{\rm eff}$ for each dwarf were computed assuming a mean age of 3~Gyr for the 
solar neighborhood (\cite{dah02}).

\begin{figure}[!b]
  \begin{center}
    \epsfig{file=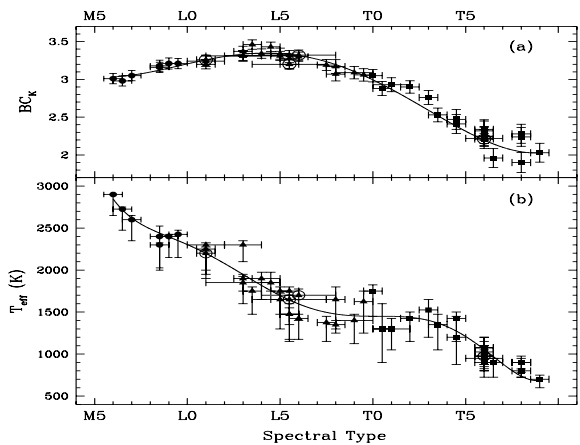, height=6cm, width=8cm}
  \end{center}
\caption{Diagrams of {\it (a)} $BC_K$ and {\it (b)} $T_{\rm eff}$ versus spectral type for ultracool
dwarfs.  The plotted values of $T_{\rm eff}$ correspond to an age of 3~Gyr, and the error bars reflect
the range of $T_{\rm eff}$ for ages 0.1--10~Gyr.  The curves are polynomial fits to the data whose
coefficients are reported by Golimowski et al.\ (2004).
\label{dgolimowskif5}}
\end{figure}

Figure~\ref{dgolimowskif5} shows the behavior of $BC_K$ and $T_{\rm eff}$ as a function of spectral
type.  $BC_K$ rises from M5 to L5 due to Wien shift and enhanced $K$ luminosity as clouds settle below 
the $K$ photosphere.  It declines after L5 because of increasing CH$_4$ absorption at $2.3~\mu$m.  The
scatter among late-T dwarfs shows the sensitivity of H$_2$ absorption to gravity variations.  
$T_{\rm eff}$
declines steeply and monotonically for types M6--L7 and T4--T9, but is nearly constant ($\sim~1450$~K)
over the L--T transition.  This constancy indicates that the condensate clouds settle, thin, or 
disintegrate over a small range of $T_{\rm eff}$.  Thus, the presently defined L and T spectral
classes are not proxies for a uniformly decreasing $T_{\rm eff}$ scale.  We compute $T_{\rm eff}
= 600$--750~K for the T9 dwarf 2MASS~0415-0935, which supplants Gl~570D as the coolest known brown 
dwarf.

\section{Comparison with Precipitating Cloud Models}

\begin{figure}[!b]
  \begin{center}
    \epsfig{file=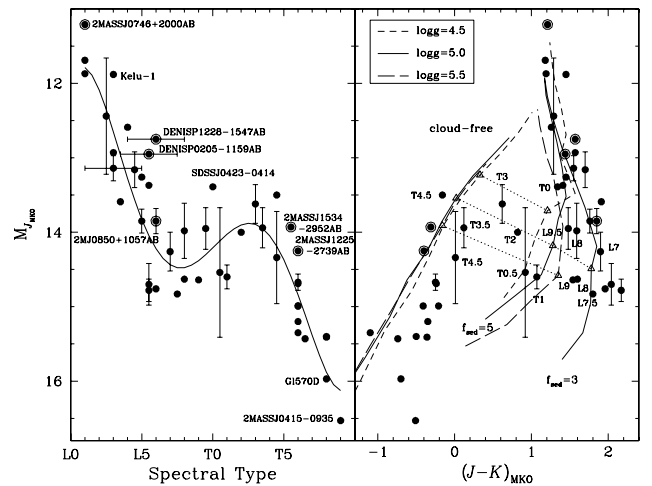, height=5.5cm, angle=-90, width=9cm}
  \end{center}
\caption{Plots of $M_J$ versus spectral type and $J$--$K$.  Error bars are shown where the uncertainties 
in the distance modulus and spectral type are $\geq 0.1$~mag and $> 1$ subclass, respectively.  Known
binaries are encircled.  Model sequences are shown in the right panel for different values of $f_{\rm sed}$
and log~$g$.
\label{dgolimowskif6}}
\end{figure}

We compare our absolute photometry with the magnitudes and colors of L and T dwarfs predicted by the 
precipitating cloud models of \cite*{ack01} and \cite*{mar02}.  The models are parametrized by 
$f_{\rm sed}$, which describes the sedimentation efficiency relative to the upward transport of 
condensates by convection.  Figure~\ref{dgolimowskif6} shows $M_J$ as functions of spectral type and
$J$--$K$ color.  A polynomial fit to the data in the left panel emphasizes the brightening $J$
luminosity between types T0 and T4.  This ``early-T'' hump (\cite{tin03}) is probably caused by 
the thinning, settling, and/or break-up of the cloud deck, which allows $J$ flux to emerge from 
deeper layers of the atmosphere.

Overplotted in the right panel of Figure~\ref{dgolimowskif6} are model sequences for $f_{\rm sed} = 3$,
5, and $\infty$ (no clouds) and different gravities (log~$g$).  The $f_{\rm sed} = 3$ models match
the L dwarf data well, but do not turn fast enough to match the T dwarf data, which are better 
matched by the cloudless models.  The dotted lines connect the $T_{\rm eff} = 1300$~K points 
{\it (triangles)} on each model sequence.  These lines bound the L--T transition, which suggests
that holes in the cloud deck develop rapidly over a narrow range of $T_{\rm eff}$.  This notion
is supported by the nearly constant $T_{\rm eff} \approx 1450$~K measured bolometrically for
types L7--T4 (Figure~\ref{dgolimowskif5}).

\begin{figure}[!t]
  \begin{center}
    \epsfig{file=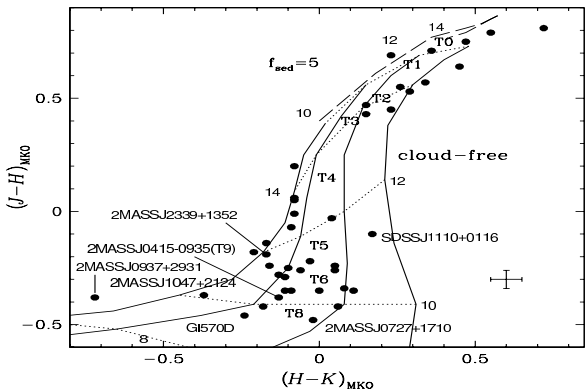, height=5.5cm, width=8cm}
  \end{center}
\caption{Plots of $J$--$H$ vs.\ $H$--$K$ for T dwarfs.  Color sequences are shown for $f_{\rm sed} = 5$,
log~$g= 5.0$ {\it (dashed curve)} and cloudless, log~$g=5.5$, 5.0, 4.5, and 4.0 {\it (solid curves, left
to right)} models.  $T_{\rm eff}$ is marked in units of 100~K; dotted lines represent constant 
$T_{\rm eff}$.  A typical error bar is shown.
\label{dgolimowskif7}}
\end{figure}

Figure~\ref{dgolimowskif7} is a $J$--$H$ versus $H$--$K$ diagram for T dwarfs overlayed with model sequences
for various combinations of $f_{\rm sed}$ and gravity.  Because the radii of brown dwarfs older than
200~Myr vary by $< 30$\% (\cite{mar96}; \cite{bur01}), gravity is tightly correlated with mass.  Thus
the solid curves in Figure~\ref{dgolimowskif7} are effectively evolutionary tracks of brown dwarfs with 
masses of {\it (left to right)} 75, 35, 15, and $10~M_{\rm Jup}$.  Consequently, the T5.5 dwarf 
SDSS~J1110+0116 may be a 10--$15~M_{\rm Jup}$ brown dwarf.  As stated in \S1, the $K$ fluxes of T dwarfs
are highly sensitive to gravity-dependent, pressure-induced H$_2$ absorption.  Thus, $H$--$K$ is a good 
indicator of mass, such that bluer $H$--$K$ reflects higher mass for a given T subtype.  

\section{Effects of Nonequilibrium Chemistry}

The model atmospheres of \cite*{mar02} assume all species are in thermochemical equilibrium.  However,
the $M$-band spectrum of the archetypal T dwarf Gl~229B shows absorption at $4.7~\mu$m by overabundant
CO dredged upward from hotter regions by convective mixing (\cite{nol97}; \cite{opp98}).  The same
phenomenon should occur in other late-T dwarfs.  

\begin{figure}[!t]
  \begin{center}
    \epsfig{file=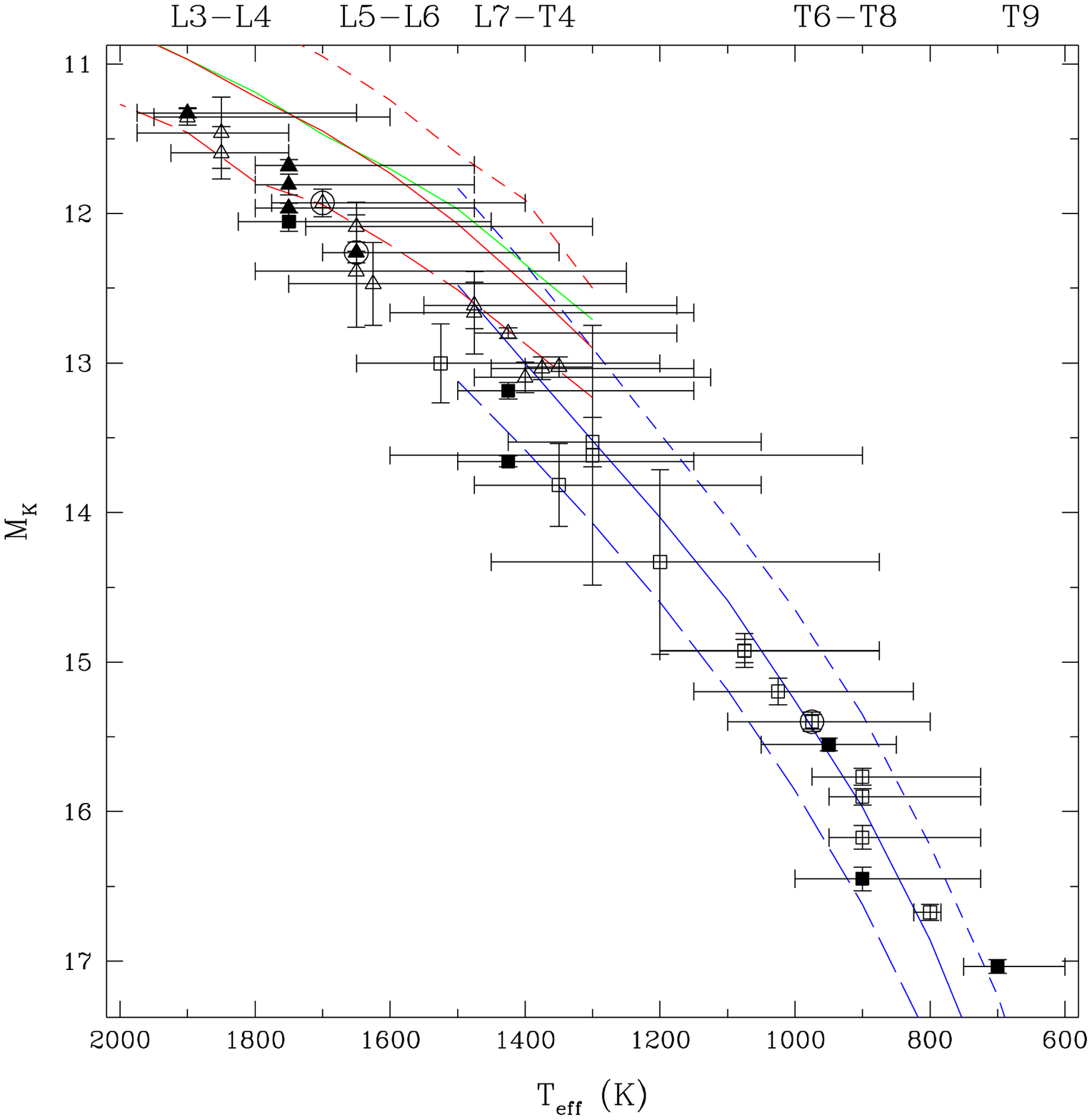, height=4.0cm, width=8cm}
    \epsfig{file=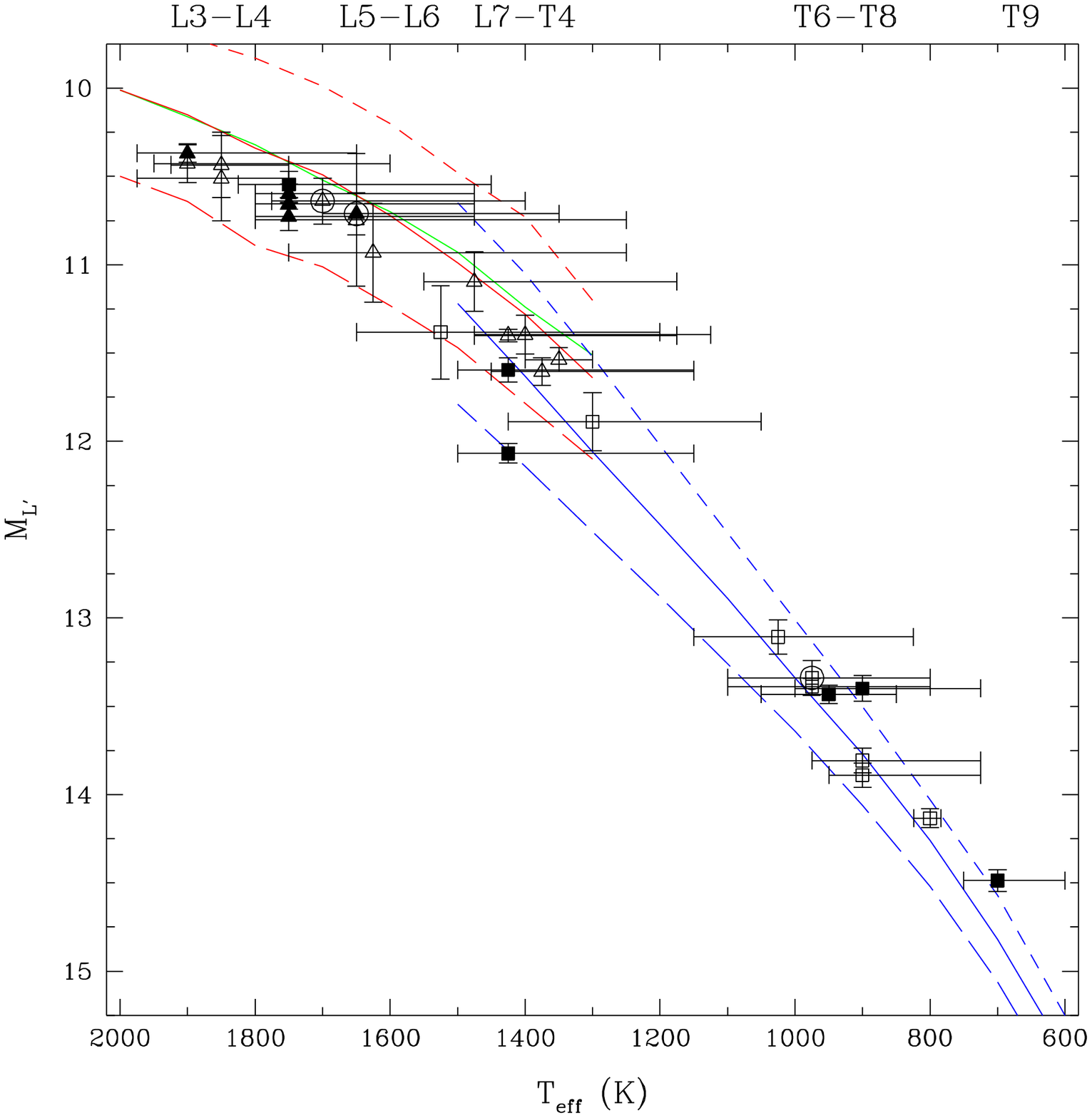, height=4.0cm, width=8cm}
    \epsfig{file=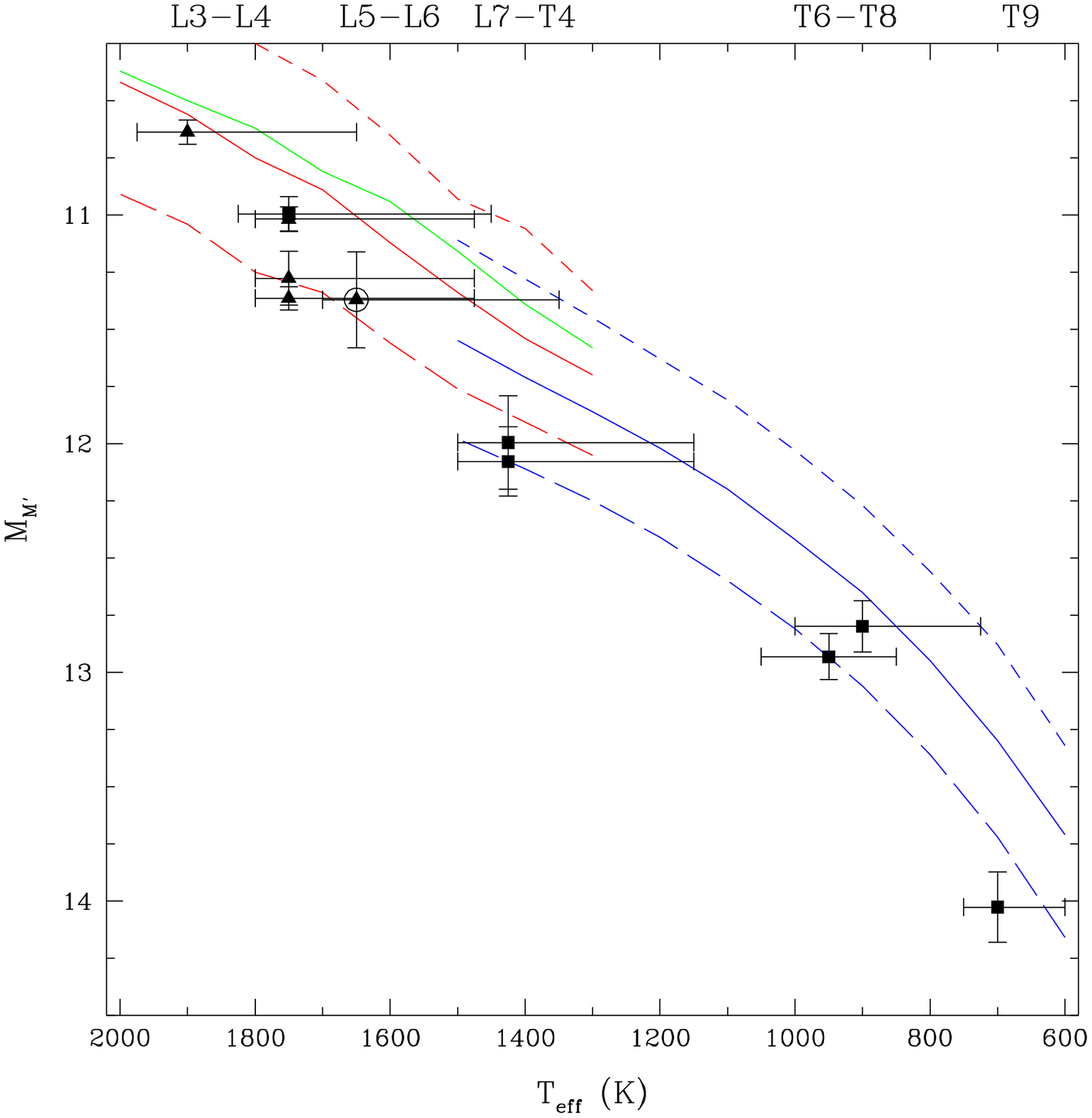, height=4.0cm, width=8cm}
  \end{center}
\caption{Plots of $M_K$, $M_{L'}$, and $M_{M'}$ vs.\ $T_{\rm eff}$ for L and T dwarfs, overlayed with
predicted sequences from equilibrium models with $f_{\rm sed} = 3$ {\it (green)}, $f_{\rm sed} = 5$ 
{\it (red)}, and cloud-free {\it (blue)} atmospheres and surface gravities of log~$g = 4.5$ {\it (short
dash)}, 5.0 {\it (solid)}, and 5.5 {\it (long dash)}.
\label{dgolimowskif8}}
\end{figure}

Figure~\ref{dgolimowskif8} compares the cloudy and cloud-free models of \cite*{mar02} with our measurements 
of $M_K$, $M_{L'}$, and $M_{M'}$ versus $T_{\rm eff}$ for L and T dwarfs.  The log~$g = 5.0$--5.5 cloudy
models match $M_K$, $M_{L'}$, and $M_{M'}$ well for L and early-T dwarfs.  The log~$g = 4.5$--5.0 
cloud-free models generally match $M_K$ and $M_{L'}$ for late-T dwarfs, but they overestimate the 
$M'$ luminosities of these dwarfs by factors of 1.5 to 2.5.  We attribute these $M'$ flux deficits to
nonequilibrium abundances of CO, as shown by the vertical-mixing models of \cite*{sau03}.  These 
overpredicted $M'$ fluxes affect the expected sensitivities of $5~\mu$m searches for even cooler,
``infra-T'' dwarfs, as are planned with the {\it Spitzer Space Telescope.}

\begin{acknowledgements}
We thank Didier Saumon for computing the model magnitudes shown in 
Figures~\ref{dgolimowskif6}--\ref{dgolimowskif8}.  This work is more completely described by 
\cite*{kna04} and \cite*{gol04}.  Tables of our photometry and spectra are available
at \linebreak http://www.jach.hawaii.edu/$\sim$skl/LTdata.html.
\end{acknowledgements}


\begin{thebibliography}{}

\bibitem[\protect\astroncite{Ackerman \& Marley}{2001}]{ack01}
Ackerman, A.~S., \& Marley, M. S. 2001, ApJ, 556, 872


\bibitem[\protect\astroncite{Baraffe et~al.}{1998}]{bar98}
Baraffe, I., et al. 1998, A\&A, 337, 403

\bibitem[\protect\astroncite{Burgasser et~al.}{2002}]{bur02}
Burgasser, A.~J., et al. 2002, ApJ, 564, 421


\bibitem[\protect\astroncite{Burrows et~al.}{1997}]{bur97}
Burrows, A., Marley, M., et al. 1997, ApJ, 491, 856


\bibitem[\protect\astroncite{Burrows et~al.}{2001}]{bur01}
Burrows, A., et al. 2001, Rev.\ Mod.\ Phys., 73, 719


\bibitem[\protect\astroncite{Chabrier et~al.}{2000}]{cha00}
Chabrier, G., et al. 2000, ApJ, 542, 464

\bibitem[\protect\astroncite{Dahn et~al.}{2002}]{dah02}
Dahn, C.~C., et al. 2002, AJ, 124, 1170

\bibitem[\protect\astroncite{Geballe et~al.}{2002}]{geb02}
Geballe, T.~R., et al. 2002, ApJ, 564, 466

\bibitem[\protect\astroncite{Golimowski et~al.}{2004}]{gol04}
Golimowski, D.~A., et al. 2004, AJ, 127, 3516

\bibitem[\protect\astroncite{Knapp et~al.}{2004}]{kna04}
Knapp, G.~R., et al. 2004, AJ, 127, 3553


\bibitem[\protect\astroncite{Marley et al.}{1996}]{mar96}
Marley, M. S., et al. 1996, Science, 272, 1919


\bibitem[\protect\astroncite{Marley et al.}{2002}]{mar02}
Marley, M. S., et al. 2002, ApJ, 568, 335

\bibitem[\protect\astroncite{Mart\'{\i}n \& Zapaterio Osorio}{2003}]{mar03}
Mart\'{\i}n, E.~L., \& Zapaterio Osorio, M.~R. 2003, ApJ, 593, L113

\bibitem[\protect\astroncite{Noll, Geballe, \& Marley}{1997}]{nol97}
Noll, K.~S., Geballe, T.~R., \& Marley, M.~S. 1997, ApJ, 489,~L87


\bibitem[\protect\astroncite{Oppenheimer et al.}{1998}]{opp98}
Oppenheimer, B.~R., et al. 1998, ApJ, 502, 932


\bibitem[\protect\astroncite{Saumon et al.}{2003}]{sau03}
Saumon, D., et al. 2003, in IAU Symp. 211, Brown Dwarfs, ed.\ E.~Mart\'{\i}n (San Francisco: ASP), 345

\bibitem[\protect\astroncite{Tinney, Burgasser, \& Kirkpatrick}{2003}]{tin03}
Tinney, C., Burgasser, A., \& Kirkpatrick, J. 2003, AJ, 126,~975

\bibitem[\protect\astroncite{Tsuji}{2002}]{tsu02}
Tsuji, T. 2002, ApJ, 575, 264

\bibitem[\protect\astroncite{Vrba et~al.}{2004}]{vrb04}
Vrba, F.~J., et al. 2004, AJ, 127, 2948

\end{thebibliography}
\end{document}